\documentclass{article}
\usepackage{spconf,amsmath,graphicx,hyperref,amssymb}
\usepackage{todonotes}
\usepackage{multirow}
\title{Fine-Grained Frame Modeling in Multi-head Self-Attention for Speech Deepfake  Detection}
\name{Phuong Tuan Dat$^{1,*}$ \qquad Duc-Tuan Truong$^{2,*}$ \qquad Long-Vu Hoang$^{1}$ \qquad \textit{Nguyen Thi Thu Trang}$^{1,\dagger}$\thanks{$^*$Co-first author. $^\dagger$Corresponding author.}}

\address{$^1$ School of Information and Communication Technology,\\Hanoi University of Science and Technology, Vietnam\\$^2$ College of Computing and Data Science, Nanyang Technological University, Singapore\\}
\begin{document}
% \ninept
%
\maketitle
\begin{abstract}
Transformer-based models have shown strong performance in speech deepfake detection, largely due to the effectiveness of the multi-head self-attention (MHSA) mechanism. MHSA provides frame-level attention scores, which are particularly valuable because deepfake artifacts often occur in small, localized regions along the temporal dimension of speech. This makes fine-grained frame modeling essential for accurately detecting subtle spoofing cues. In this work, we propose fine-grained frame modeling (FGFM) for MHSA-based speech deepfake detection, where the most informative frames are first selected through a multi-head voting (MHV) module. These selected frames are then refined via a cross-layer refinement (CLR) module to enhance the model’s ability to learn subtle spoofing cues. Experimental results demonstrate that our method outperforms the baseline model and achieves Equal Error Rate (EER) of 0.90\%, 1.88\%, and 6.64\% on the LA21, DF21, and ITW datasets, respectively. These consistent improvements across multiple benchmarks highlight the effectiveness of our fine-grained modeling for robust speech deepfake detection.
\end{abstract}
\section{Introduction}
\label{sec:intro}

The rise of deepfake speech, enabled by advances in text-to-speech (TTS) \cite{article} and voice conversion (VC) \cite{singing_challenge}, poses serious threats to biometric security systems in banking, telecommunications, and access control \cite{8299800}, as well as broader risks like misinformation and social engineering attacks \cite{7400997, WU2015130}. To address these challenges, initiatives such as the ASVspoof series (2015–2021) \cite{10155166} and the recent VSASV 2024 challenge \cite{hoang2024vsasv} have driven the development of synthetic speech detection (SSD) methods \cite{truong24_asvspoof, Nes2Net, phuong25_interspeech}. Among these, transformer-based architectures, especially those leveraging pre-trained self-supervised learning (SSL) models like wav2vec 2.0 \cite{baevski2020wav2vec20frameworkselfsupervised} and WavLM \cite{9814838}, have emerged as the current state-of-the-art. This is largely due to the multi-head self-attention (MHSA) mechanism, which captures frame-level dependencies, with each attention head exhibiting distinct sensitivity to acoustic patterns \cite{audhkhasi22_interspeech}, making these models well-suited for detecting artifacts in synthetic speech.

Despite the strong modeling capacity, MHSA treat attention outputs as globally aggregated features and overlook the fine-grained temporal dynamics that are crucial for synthetic speech detection. In vanilla MHSA, attention is distributed across the entire sequence, but without explicit mechanisms to highlight or prioritize localized anomalies, subtle spoofing cues may be diluted or ignored. Prior studies have shown that synthetic artifacts can manifest in short, temporally sparse regions, such as unnatural transitions \cite{boundary} or specific phonemes \cite{phoneme}. These localized anomalies can escape detection when frame-level modeling is too coarse or implicitly averaged across the sequence. Furthermore, recent work has demonstrated that different attention heads specialize in capturing different types of acoustic patterns \cite{audhkhasi22_interspeech}, suggesting that a head-level selection or weighting mechanism could enhance sensitivity to spoofing artifacts.

Motivated by these observations, we propose a fine-grained frame modeling (FGFM) approach for speech deepfake detection. The proposed FGFM first applies multi-head voting (MHV) to select high-attention frames from each MHSA head, identifying the most salient fine-grained frames for spoof detection. These selected frames are then aggregated across different layers via a cross-layer refinement (CLR) module and integrates this refined information to enrich the classification token. While MHV and CLR modules was originally introduced for fine-grained visual classification tasks \cite{xu2023fine}, we demonstrate its broader applicability to speech deepfake detection by showing that enhancing selective information modeling is well-suited for capturing the subtle and localized artifacts characteristic of deepfake speech. Through evaluations across multiple benchmarks, our proposed method outperforms the baseline model on ASVspoof 2021 LA, DF, and the out-of-domain In-the-Wild (ITW) dataset, with relative EER reductions of 7.2\%, 27.1\%, and 21.1\%, respectively. %over baselines and competitive state-of-the-art models. %: ASVspoof 2021 LA, DF, and the out-of-domain In-the-Wild (ITW) dataset, with relative EER reductions of 7.2\%, 27.1\%, and 21.1\%, respectively.

\begin{figure*}[h]
    \centering
    \includegraphics[width=0.7\linewidth]{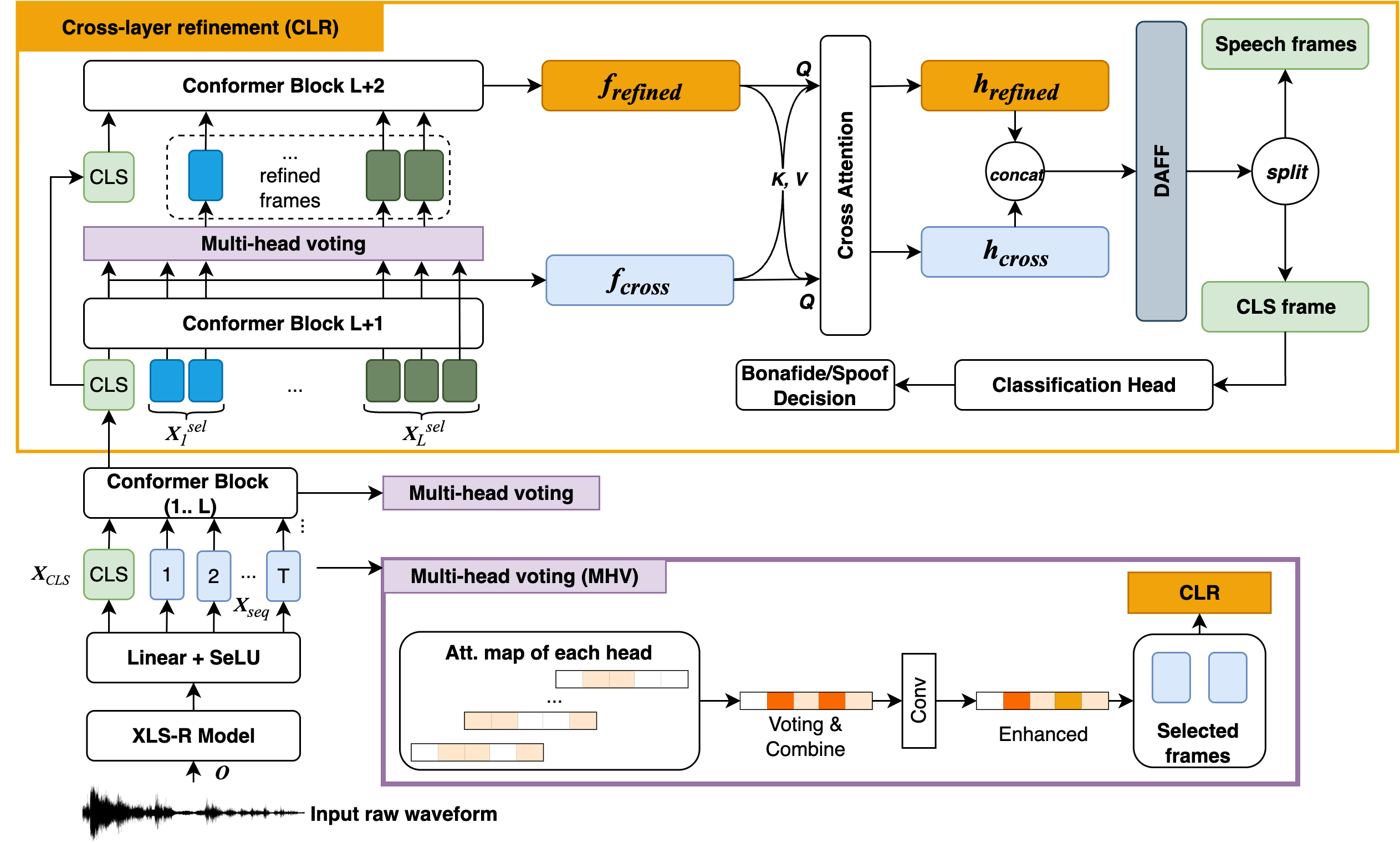}
    \caption{Overview of the proposed Fine-Grained Frame Modeling (FGFM) architecture. 
    % DAFF stands for dynamic aggregation feed forward module.
    }
    \label{fig:architecture}
\end{figure*}
\section{Methodology}

Figure~\ref{fig:architecture} provides an overview of our proposed Fine-Grained Frame Modeling (FGFM) method. We first describe the baseline XLSR-Conformer model \cite{rosello23_interspeech}, followed by the components of the proposed method: the Multi-Head Voting (MHV) and Cross-Layer Refinement (CLR) modules.
% These vision-inspired adaptations are specifically optimized for capturing synthetic speech artifacts across temporal dimensions.

\subsection{The baseline model}
\label{sub:baseline}
Our baseline is the XLSR-Conformer model \cite{truong24b_interspeech}, which combines a pre-trained XLS-R model \cite{babu2021xls} as a feature extractor (FE) with a Conformer model. Given an input signal $O$, the FE outputs $T$-frame sequent embedding, which are projected to a lower $D$-dimensional  space by a linear projection layer. A learnable classification token $X_{\text{CLS}}$ is prepended to projected input embedding $X_{seq}$, forming the encoder input $X$. This input sequence is passed through $L$ MHSA-based Conformer blocks, each produces the output representation  $X^l$. Finally, the updated classification token from the last block is passed through a classification head to determine whether the input is bonafide or spoofed. This model forms the foundation upon which we introduce our fine-grained frame modeling.

\subsection{The proposed model}

Our proposed FGFM architecture incorporates the MHV, and CLR modules, as illustrated in Figure \ref{fig:architecture}. We describe the mechanism of each module in this section.

\subsubsection{Multi-head voting (MHV) module}
The MHSA in each Conformer block divides the input into $K$ attention heads in the temporal dimension and feeds each group to each head to calculate self-attention. Each block $l$ generates $K$ attention maps $A^l = [A^1, ..., A^K]$ between the classification token and other speech frames, $A^k \in \mathbb{R}^T$. However, the attention map of each head can represent a concentration on different areas for discrimination cues \cite{vaswani2017attention}. To mitigate the effects of heads with poor classification performance and to adaptively learn to select valuable frames from each Conformer block, we propose to use the MHV module to treat each head in MHSA as a weak learner for a bagging algorithm to select valuable frames for effective discrimination.

Specifically, for all Conformer block from  $1$ to $L$, MHV learns to vote to select $v$ frames per attention head, based on each attention map in $A^l$. The selected frames will be flagged with a value of $1$ in a binary score map $M^k$, and $0$ otherwise. Finally, we combine the score map across $K$ attention heads with a summation, $M' = \sum_{k=0}^KM^k$. To further enhance the score map, we apply a convolution operation with a 1-D Gaussian-like kernel to yield the refined map $M^*=M' *\mathbf{G}$, where $\mathbf{G}$ is presented in Equation \ref{eq:matrix}. From the MHV module at each Conformer block $l$, $v$ frames with the highest scores in $M^*$ are selected, whose representations are denoted as $X_l^{sel} = [X^{sel}_{l,1}, ..., X^{sel}_{l,v}]$.
\begin{equation}
\label{eq:matrix}
    \mathbf{G} = \begin{bmatrix}
1 & 2 & 3 & 4 & 3 & 2 & 1
\end{bmatrix}
\end{equation}

\subsubsection{Cross-layer refinement (CLR) module.} While MHV extracts valuable frames within individual blocks, effective spoof detection also requires exploiting cross-layer information. To this end, we introduce the CLR module. At the $(L+1)$-th Conformer block, the input is formed by concatenating the classification token with all previously selected frames: $X_L^{in}=[X_{L}^\text{CLS}; X_1^{sel};...;X_{L}^{sel}]$. The output frame sequence, denoted $f_{\text{cross}} = [X^{\text{CLS}}_{L+1}; X_{L+1}^{sel}]$, captures global discriminative cues aggregated across layers.

% To obtain the refined feature for enriched discrimination cues, we employ an additional $(L+2)$-th Conformer block to ingest the refined frames from $X_{L+1}$ with the MHV module. Besides, to prevent the refined feature from the effect of noise, we use the classification token from the $L$-th rather than the $(L+1)$-th block to concatenate with the refined frames. Finally, the input of the additional $(L+2)$-th Conformer block is $X_{L+2}^{in} = [X_{\text{CLS}}^{L}; X_{L+1}^{sel}]$. The refined feature is the classification token from the output features of the $(L+2)$-th Conformer block, $f_{cross} \equiv X_{\text{CLS}}^{L+2}$, which uses prior prediction results to guide the final classification.

To obtain refined features, we apply the MHV module again on the $(L+1)$-th block’s outputs, selecting frames from $X_{L+1}^{sel}$. These, together with the classification token from block $L$, are passed to an additional $(L+2)$-th Conformer block: $X_{L+2}^{in} = [X_{L}^\text{CLS}; X_{L+1}^{sel}]$. The output sequence from this block is denoted as the refined feature $f_{\text{refined}} = [X^{\text{CLS}}_{L+2}; X^{sel}_{L+2}]$. To exchange information between $f_{\text{cross}}$ and $f_{\text{refined}}$, we adopt two parallel cross-attention operations, where the inputs are the vectors of
query $(Q_{\text{cross}}, Q_{\text{refined}})$ key $(K_{\text{cross}}, K_{\text{refined}})$, and value $(V_{\text{cross}}, V_{\text{refined}})$ from cross-layer and refined feature, $f_{\text{cross}}$ and $f_{\text{refined}}$ respectively, projected by a linear layer:

\begin{align*}
% \label{equ:attention1}
h_{\text{cross}} = \text{softmax}\left( \frac{Q_{\text{cross}} \cdot K_{\text{refined}}^T}{\sqrt{D}} \right) V_{\text{refined}} + f_{\text{cross}}\\
h_{\text{refined}} = \text{softmax}\left( \frac{Q_{\text{refined}} \cdot K_{\text{cross}}^T}{\sqrt{D}} \right) V_{\text{cross}} + f_{\text{refined}}
\end{align*}

The concatenated output $[h_{\text{cross}}; h_{\text{refined}}]$ is then processed through a lightweight DAFF block \cite{lu2022bridging}, which acts as a feed-forward aggregator to strengthen cross-correlations and enhance the final classification token. The enriched classification token, which integrates information from both intra-layer (MHV) and cross-layer (CLR$+$DAFF) processing, is finally passed to the classification head for the binary bonafide/spoof decision. This design enables FGFM to capture fine-grained artifacts from synthesized speech.%, achieving robustness even against unseen spoofing techniques.

\section{Experiments}

\subsection{Datasets and experimental setup}
All models are trained on the ASVspoof 2019 LA training set. We evaluate performance on three benchmarks: ASVspoof 2021 LA (21LA), ASVspoof 2021 DF (21DF), and the In-the-Wild (ITW) dataset \cite{muller22_interspeech}. While ASVspoof 2021 LA and DF speech data is distorted by various codec and compression variations, ITW dataset presents a different challenge due to its coverage of previously unseen device/channel mismatches, and naturalistic variability. We report performance using Equal Error Rate (EER) as the main metric.

We follow the same model configuration as our baseline Conformer model \cite{truong24b_interspeech}, with the addition of two Conformer encoder blocks in the CLR module. For the MHV module, we set the number of fine-grained frames selected by each attention head in each encoder block to $v = 24$, a value determined empirically. The training setup remains consistent with that of the baseline. All experiments are conducted on a single NVIDIA A40 GPU. 
\subsection{Comparison with SOTA models}
The performance of the proposed model and the comparison with recent SOTA models are reported in Table \ref{tab:baseline-results}.
The proposed XLSR-MHV, achieves SOTA performance across all three benchmarks in terms of EER, outperforming recent systems. Compared to the baseline XLSR-Conformer, it achieves relative EER improvements of 7.2\% on 21LA, 27.1\% on 21DF, and 21.1\% on the ITW dataset. While performing on par with XLSR-Mamba on the 21DF set, the proposed model surpasses all others on the out-of-domain ITW set with a leading 6.64\% EER, demonstrating strong generalization and robustness to unseen spoofing conditions. In addition, we replace the Conformer blocks with Transformer blocks to validate the robustness of our proposed modules. It can be seen that the proposed methodology brings stable improvement for both Conformer-based and Transformer-based model, improving the result by 12.2\% relatively on average. This suggests that our methodology is beneficial and generalizable to various MHSA-based architectures.
\begin{table}[t!]
\centering
% \usepackage{multirow}
% \usepackage[normalem]{ulem}
    \begin{tabular}{lccc}
    \hline
    \multirow{2}{*}{\textbf{Model}} & \multicolumn{3}{c}{\textbf{EER (\%)}}           \\ \cline{2-4} 
                                    & \textbf{21LA} & \textbf{21DF} & \textbf{ITW}  \\ \hline
    XLSR-Conformer+TCM \cite{truong24b_interspeech}$\dagger$              & 1.18           & 2.25           & 7.79          \\
    XLSR-AASIST \cite{tak2022automaticspeakerverificationspoofing}                    & 1.00           & 3.69           & 10.46         \\
    XLSR-AASIST2 \cite{10448049}                   & 1.61           & 2.77           & -             \\
    WavLM-MFA \cite{10447923}                      & 5.08           & 2.56           & -             \\
    XLSR-SLS \cite{zhang2024audio}                       & 5.08           &  \underline{1.92}           & 7.46          \\
    XLSR-MoE \cite{10888990}                       & 2.96           & 2.54           & 9.17          \\
    XLSR-Mamba \cite{10909468}                     & \underline{0.93}     & \textbf{1.88}  &  6.71    \\ \hline
    XLSR-Transformer                & 1.96             & 2.43             & 6.59             \\
    \quad + FGFM (ours)    & 1.82             & 2.37             & \textbf{6.31}            \\ \hline
    XLSR-Conformer \cite{rosello23_interspeech}$\dagger$                  & 0.97           & 2.58           & 8.42          \\
    \quad + FGFM (ours)      & \textbf{0.90}  & \textbf{1.88}  & \underline{6.64} \\ \hline
    \end{tabular}
  \caption{Performance comparison with the state-of-the-art systems on the ASVspoof 2021 eval set and ITW dataset ($\dagger$ denotes reproduced results, \textbf{bold} denotes the best result, \underline{underline} denotes the second-best result, and `-' denotes the results are unavailable).}
  \label{tab:baseline-results}
\end{table}

\subsection{Ablation studies}

To better understand the contribution of each component in our proposed framework, we conduct a series of ablation studies based on the XLSR-Conformer baseline. The results are summarized in Table~\ref{tab:ablation}. We first compare the baseline system using an increased number of Conformer blocks ($L = 6$) to match the depth of our proposed model, which includes four original Conformer blocks and two additional blocks in the CLR module. The results show that simply increasing the number of Conformer blocks in the baseline does not achieve comparable performance as our method, indicating that our improvements are not due to increased model capacity. Furthermore, by systematically removing or modifying individual components, we evaluate their impact on overall performance to highlight the effectiveness of each design choice.
\subsubsection{The effectiveness of each proposed module}
First, we remove the DAFF module, letting the cross-layer feature $f_{\text{cross}}$ to be passed directly to the classification head. This modification results in relative performance drops of 8.2\%, 5.1\%, and 2.7\% on the 21LA, 21DF, and ITW datasets, respectively. However, the model still performs comparably to the XLSR-Conformer baseline, and notably outperforms it on the ITW dataset. These results suggest that the DAFF module improves performance by effectively integrating information from fine-grained speech frames into the classification token.  Secondly, we remove the enhancement operation with Gaussian-kernel convolution in the MHV module. This leads to relative performance reductions of 15.6\%, 8.0\%, and 5.1\% on the 21LA, 21DF, and ITW test sets, respectively. These findings confirm that the enhancement step is beneficial for refining the voted attention map.

\begin{table}[t]
\centering
    \begin{tabular}{lccc}
    \hline
    \multirow{2}{*}{\textbf{Model}} & \multicolumn{3}{c}{\textbf{EER (\%)}}           \\ \cline{2-4} 
                                    & \textbf{21LA} & \textbf{21DF} & \textbf{ITW}  \\ \hline
XLSR-Conformer (baseline) & 0.97  & 2.58 & 8.42 \\
XLSR-Conformer ($L=6$) & 2.09 & 2.68 & 9.79 \\ \hline
XLSR-Conformer-FGFM                  & \textbf{0.90}  & \textbf{1.88} & \textbf{6.64} \\
\hspace{5mm}-w/o DAFF     & 0.98  & 1.98 & 6.82 \\
\hspace{5mm}-w/o enhancement in MHV & 1.04  & 2.03 & 6.98 \\ \hline
\end{tabular}
\caption{The performance of XLSR-MHV and baseline model with and without the DAFF and enhancement modules.}
\label{tab:ablation}
\end{table}

\begin{table}[t]
\centering
% \resizebox{\columnwidth}{!}{%
\begin{tabular}{lccc}
\hline
\multirow{2}{*}{\textbf{Vote per head}} & \multicolumn{3}{c}{\textbf{EER (\%)}}        \\ \cline{2-4} 
                                        & \textbf{21LA} & \textbf{21DF} & \textbf{ITW} \\ \hline
$v = 16$                                      & 1.34          & 2.04          & 7.04         \\
$v = 24$                                      & \textbf{0.98}          & \textbf{1.98}          & 6.82         \\
$v = 32$                                      & 1.69          & 2.27          & 6.73         \\
$v = 40$                                      & 1.71          & 2.84          & \textbf{6.66}         \\ \hline
\end{tabular}
% }
\caption{Effect of the number of vote per head $v$ in MHV.}
\label{tab:tune_vote_per_head}
\end{table}

\subsubsection{The number of voted frames per each head in MHV module}

Table~\ref{tab:tune_vote_per_head} presents an ablation study on the effect of varying the number of votes per head in the MHV module across three evaluation datasets. Performance improves as the number of votes increases from 16 to 24, achieving optimal EERs of 0.98\%, 1.98\%, and 6.82\% on LA21, DF21, and ITW, respectively. However, further increasing to 32 and 40 votes results in performance degradation across all datasets. This drop is likely due to the inclusion of less informative frames, such as those representing silence or non-speech regions, which introduce noise and reduce discriminative power. These results indicate that selecting 24 votes per head strikes an effective balance for 4-second audio segments, capturing key speech frames while excluding irrelevant ones.

\subsubsection{Key features that models focus on in speech}
\begin{figure}[h]
    \centering
    \includegraphics[width=\linewidth]{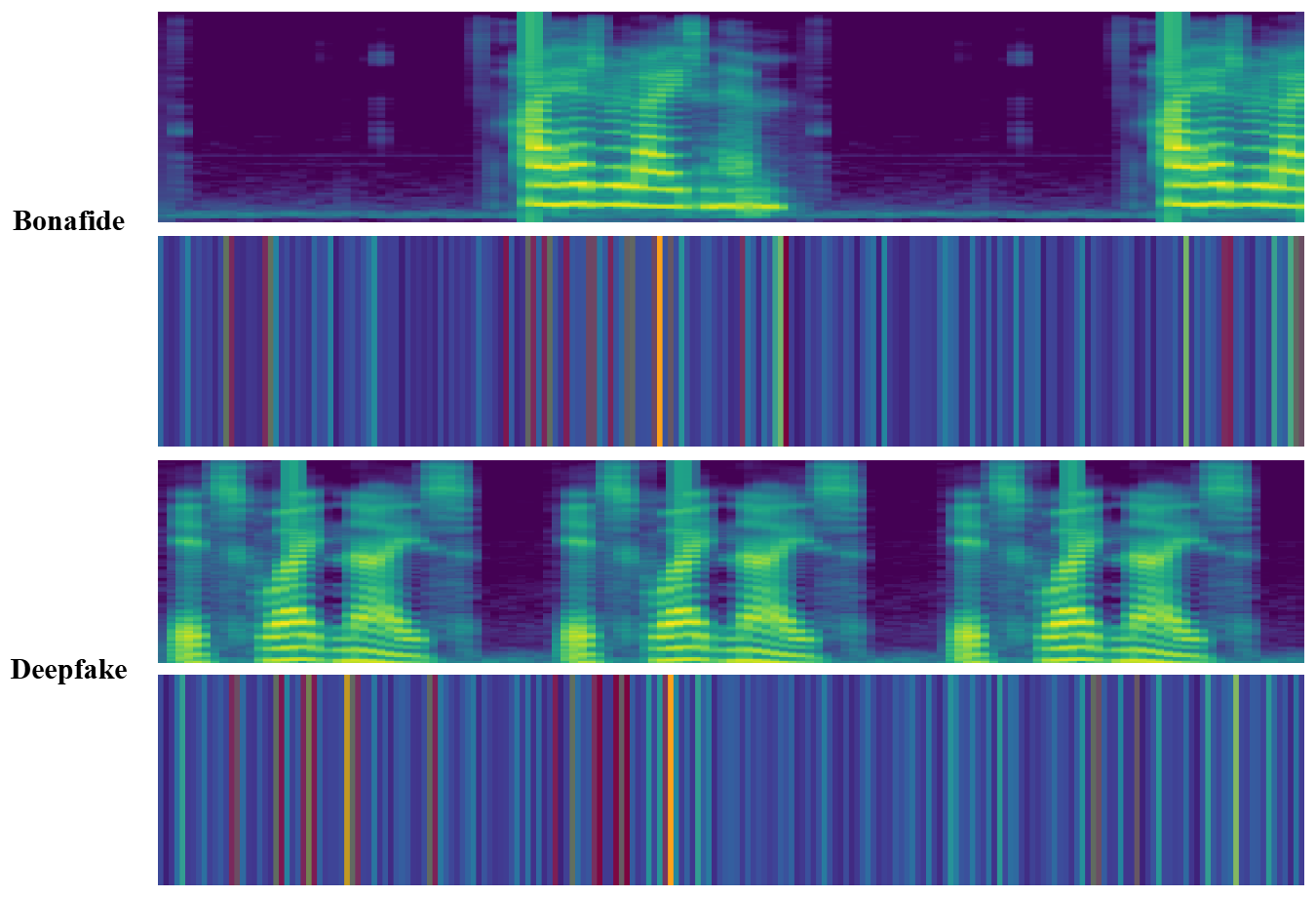}
    \caption{Visualization of selected frames (red vertical lines) in MHV, in relation to spectrograms for bonafide (top) and deepfake (bottom) audio samples.}
    \label{fig:plot_attention}
\end{figure}
Figure \ref{fig:plot_attention} compares attention maps from the final Conformer block with spectrograms of bonafide and deepfake audio. The red lines mark frames selected by our MHV module, showing that it consistently attends to speech-rich regions while avoiding silence or low-information segments. In both audio types, the selected frames align with areas of high-energy speech-rich regions. This targeted focus is crucial, as speech signals often include silent or uninformative intervals that can weaken classification if treated equally.

% The visualization confirms that adapting MHV from vision to speech effectively handles the uneven distribution of discriminative features across time. By emphasizing information-rich segments, MHV boosts the signal-to-noise ratio in feature representations, leading to the performance gains observed in our results. This highlights how vision-inspired modules can be successfully re-purposed for audio when tailored to its unique characteristics.

\section{Conclusion}
This paper presented a fine-grained frame modeling approach for speech deepfake detection that enhances the ability of MHSA-based models to capture subtle and localized artifacts. Our framework integrates a multi-head voting mechanism to select salient frames, a cross-layer refinement module to aggregate hierarchical information, and a dynamic aggregation feed-forward (DAFF) module to enrich the classification token with contextually important features. Extensive experiments on ASVspoof 2021 LA, DF, and the In-the-Wild datasets demonstrate that our method consistently outperforms strong baselines and competitive state-of-the-art systems. These results highlight the effectiveness of selective information modeling for improving the robustness of speech deepfake detection models.

\section{Acknowledgement}
This research was funded by the Ministry of Education and Training of Vietnam under project code CT2025.EA.BKA.04.
% \vfill\pagebreak


\begingroup
\footnotesize
\bibliographystyle{IEEEbib}
\bibliography{strings,refs}
\endgroup

\end{document}